\documentclass[11pt]{amsart}
\usepackage{geometry}                
\usepackage{graphicx}
\usepackage{amssymb}
\usepackage{epstopdf}
\usepackage{listings}
\usepackage{xcolor}
\usepackage{hyperref}
\usepackage{courier}

\definecolor{darkgreen}{RGB}{0,102,0}
\definecolor{darkblue}{RGB}{0,0,102}
\definecolor{darkgray}{RGB}{96,96,96}
\definecolor{teal}{RGB}{0,139,139}
\definecolor{soothingred}{RGB}{255,0,107}

\usepackage[T1]{fontenc}

\lstdefinelanguage{jxl}
{
  morekeywords={
    import, if, else, while, for, file, as, until , return , continue, break, new , atomic, def , size , empty, 
    read, write, println , send, random, select , where , join, list, set, mset, dict, array, sub,
    lfold, rfold, index, rindex , hash , true, false , null , matrix , json, xml , error , tokens , 
    enum , lines , partition , sorta, sortd , num, heap, db ,
    isa , type , bye, test , thread, system , minmax, sum , load , clock, shuffle ,
    int, INT, double, DEC, time,  str , fopen , exists , 
    \$, @ , \_ , \$\$   , @ARGS , \#
  },
  sensitive=true, 
  morecomment=[l]{//}, 
  morecomment=[l]{\#\#}, 
  morecomment=[s]{/*}{*/}, 
  morestring=[b]", 
  morestring=[b]', 
  morestring=[b]`, 
}
\usepackage[T1]{fontenc}
\lstdefinestyle{ZoomBAStyle}{
  language=jxl,
  basicstyle=\fontsize{9}{10}\selectfont\ttfamily, 
  commentstyle=\itshape\color{darkgray}, 
  keywordstyle=\bfseries, 
  stringstyle=\itshape, 
  extendedchars=true, 
  tabsize=2, 
  showstringspaces=false, 
  breaklines=true, 
  frame=lines, 
  framesep=5pt, 
  xleftmargin=5pt,
  xrightmargin=5pt
}

\title{ZoomBA \\ A declarative Language for Rapid Business Development}
\author{Nabarun Mondal, Jatin Puri, Mrunal Lohia \\ BayesTree Inc.}

\begin{document}
\maketitle

\begin{abstract}
The motivation for ZoomBA are domain specific languages (DSL) like VERILOG \cite{verilog}, VHDL \cite{vhdl}, Spice \cite{spice}.
DSL for Software Integration and testing is not a new idea, many commercial tools
like SAP's ABAP \cite{abap}, Silk Suite \cite{silk} use them, while Selenese, the DSL for Selenium IDE \cite{selenese}
is open source. ZoomBA is a functionally motivated, embeddable, Turing Complete \cite{turing} \cite{church} micro language. 
It's goal is to expose existing runtime echo systems in a declarative fashion for the purpose of 
System Integration and software validation \cite{dta1}. 
By design ZoomBA script size is meagre compared to Python or Scala for business automation problems. 
Bayestree uses ZoomBA for system integration, data adapter and manipulation as well as automated testing purposes.
\end{abstract}

\begin{section}{Introducing ZoomBA}

\begin{subsection}{Introductory Remarks}
Tenet of today is : \emph{Churn code out, faster, and with less issues and make end users happy}. 
That is the precise idea behind of Test Driven Development (TDD)\cite{dta1}. 
The economy of testing in a fast paced release cycle puts automatic validation into foray.
Still, given there are not many people allocated for the testing efforts overall
( some firms actually makes user do their testing ) there is an upper cut-off on what can be accommodated,
even through automation. Hence, the idea is \emph{Getting More Done, with Less}. ZoomBA sprang out of this idea,
a scripting language based on Java Virtual Machine (JVM), embeddable in any other JVM language, concise to write, and 
declarative by design, so that formal logic \cite{dp} can be encoded cleanly using ZoomBA.
Later, due to the same ``provability'' trait it was found to be a very suitable tool for System Integration purposes.
\end{subsection}

\begin{subsection}{ZoomBA at a Glance}
The syntax of ZoomBA is influenced by Python\cite{rossum}, Scala\cite{scala} and Go\cite{go}. 
ZoomBA keywords are pretty standard, because all of them are borrowed from mainstream languages \cite{sicp}:
\emph{
\begin{enumerate}
\item{ \textbf{Control Flow} : if, else, for, while, break, continue, return. }
\item{ \textbf{Definitions} : def, is, import, as, where, isa, type.}
\item{ \textbf{Literals} : 'string' , "string" , true, false, null. }
\item{ \textbf{Object Creation} : new. }
\end{enumerate}
}

Basic data structures are \emph{array, list, set, dict, heap.}
ZoomBA is a dynamically typed language. 
Elementary data types are \emph{chrono : time; integer : int, INT; float : double, DEC ; number : num ; range }.
There are utility functions by default which takes care of type conversions and type check, comparisons, and comprehensions
on collections. Most of the assignment operations automatically uses appropriate container type for a variable. 
Runtime type checks and programmatic type checks are part of the specification.
Global variables are defined using prefix of \emph{\$} : $\$global$.

\begin{lstlisting}[style=ZoomBAStyle]
x = int('42', 0) // cast 42 into int, if failed, return 0 
d = time('19470815','yyyyMMdd' ) // string to a date using format 
a = [1,2,3] // a is a fixed size item-mutable list comprise of integers  
d = { 0 : false , 1 : true } // d is a dictionary 
s = set(1,2,2,2,3) // s is a set of (1,2,3)
// f is converted into Real, no loss of precision
f = 0.100101000017181881881888188981313873444111 
// fp is a variable, containing a nameless function adding its two params
fp = def(a,b){ a + b }
/* type checking... */ 
0 isa 'int' // true 
0.42 isa 'float' // true 
time() isa 'date' // true 
[0,1] isa 'list' // true 
\end{lstlisting}

There are higher order functions available, and every function is capable of taking a function as input.
Nesting of functions is permitted. Partial functions exist, and strings can be evaluated as ZoomBA scripts.

Interaction with underlying system and network is possible via i/o operations like:
\emph{read,write,println,send} and creating a process and a thread is possible via 
\emph{system,thread} functions.

\end{subsection}

\begin{subsection}{Introducing Control Flow : FizzBuzz}
To showcase control flow, FizzBuzz is a good example. The problem statement is : 
\emph{Given a number is divisible by 3 print Fizz,
if divisible by 5 print Buzz, and if neither print the number.} The solution is as follows:

\begin{lstlisting}[style=ZoomBAStyle]
def fizz_buzz(upto){ // defining a method with a known named parameter
   // create dictionary 
   fb_hash = { 0 : 'FizzBuzz' , 3 : 'Fizz' , 5 : 'Buzz', 
               6 : 'Fizz' , 9 : 'Fizz' , 10 : 'Buzz' , 12 : 'Fizz' }
   for ( i : [1:upto+1]){ // iterate over a range             
      r = i % 15 // modulo operator
      // if r in the hash, then, else... classic ternary 
      println (  r @ fb_hash ? fb_hash[r] : i ) 
   }
}
\end{lstlisting}
\end{subsection}

\begin{subsection}{Anonymous Function as Parameter in Higher Order Functions }
To demonstrate the feature of using higher order functions,
we solve the problem of finding the largest line from a text file.

\begin{lstlisting}[style=ZoomBAStyle]
def largest_line(file_name){
   line_iter = file(file_name) // gets a lazy (iterator) of lines 
   // comparator : size of left < size of right implies left < right 
   #(min,MAX) = minmax( line_iter ) where { size($.left) < size($.right) } 
   // multiple assignment sets min to min, and MAX to MAX length string  
   println(MAX)
}
\end{lstlisting}

SQL like syntax is available in the form of \emph{join} which is 
the most general form of collection interwinding. The following code
generates all permutations of a string \emph{word}:

\begin{lstlisting}[style=ZoomBAStyle]
n = #|word| // get cardinal size of a collection  
l = [0:n].list() // get a list out of a range 
ll = list([0:n]) as { l } // get a list of list 
permutations = set() // create empty set 
join(@ARGS = ll ) /* assign all the arguments from ll */ where{
    // cast to set and compare, to remove repeat
    continue( #|set($.o)| != #|$.o| ) 
    indices = $.o // store item 
    p = str(indices,'') as { word[$.o] }
    permutations += p  
    false // do not add anything to join result
} 
\end{lstlisting}

\end{subsection}

\begin{subsection}{Reading URL, Clocking, List Comprehensions, Sorting }
Suppose we do want to test how fast certain portion 
of code runs. ZoomBA comes with a default construct to solve it. Here, 
in the below code we benchmark the load time of https://arxiv.org :

\begin{lstlisting}[style=ZoomBAStyle]
def benchmark(url, count){
   def get_time(url){ // nested function 
     #(t,o) = #clock { read(url) } // clock the read nanoseconds
     return t 
   }
   timings = list([0:count]) as {  get_time(url) } // comprehension 
   timings = sortd(timings) // sorting descending 
   i = int ( 0.1 * count ) // find 90% 
   timings[i] // return is optional keyword
}
t = benchmark( 'https://arxiv.org/' , 30 )
println(t)
\end{lstlisting}
\end{subsection}

\begin{subsection}{Import, Error Handling}
The underlying system is capable of raising errors, so while ZoomBA
does not support \emph{try...catch...finally} constructs, it is capable
of raising and catching errors.

\begin{lstlisting}[style=ZoomBAStyle]
import 'java.lang.Integer' as Int // import Integer class as alias Int
// ( output ? error ) signifies the error to be caught , call static parseInt 
#(o ? e) = Int:parseInt('The answer to everything is 42')
println(o) // is null
println(e) // ZException$Function: parseInt : For input string: ...  
\end{lstlisting}

\end{subsection}

\end{section}

\begin{section}{ZoomBA in Practice}

In this section we introduce three problems of increasing difficulty, 
discuss where the problem really lies, and explain 
formulation of such a problem using mathematical logic, 
and then we solve them declaratively using ZoomBA \cite{ZoomBA}.

\begin{subsection}{Validation of Sorting}
There is a function $S$ which takes a list of objects $l_i$ ( say integers )
and returns a sorted list $l_o$, how to test that the $S$ did the job correctly \cite{dta1}?
The problem is two faced, not only it requires to verify that the $l_o$ is ordered,
but also that $l_o$ is a permutation of $l_i$. Suppose $I = \{1,2,...,|l_o|-1 \}$.
Formally, the validation now becomes :
$$
l_o = l_i \; and \; \not \exists i \in I \; such \; that \; l_o[i-1] > l_o[i] 
$$

Here is how to solve it using ZoomBA:
\begin{lstlisting}[style=ZoomBAStyle]
def is_sorted_permutation(l_i,l_o){
    /* '==' tests permutation [1,2] == [2,1] 
    $.i defines the current item index, $.o is the current item, $.c the list 
    exists function returns true/false for a condition
    which was specified in the braces known as anonymous function block -->{ } */
    return ( l_i == l_o && 
    !exists(l_o) where { $.i > 0 && $.c[$.i -1] > $.o }  )
}
\end{lstlisting}
\end{subsection}

\begin{subsection}{Filtering}
There is a function $F$ which takes a list of objects $l$, and a predicate $P$ \cite{dt},
and generates a list: $F(l,P) = l_F$. 
Clearly $l_F \subset l$ such that $l_F = \{ x \in l \; | P(x) \;=\; True \}$.
How to verify that the function $F(P,l)$ worked correctly? 
Here is the code in ZoomBA that solves it :
\begin{lstlisting}[style=ZoomBAStyle]
def verify_applied_filter(P, l, l_F){
  //find the failure  
  failure = exists(l_F) where { !P($.o) }
  // sublist check is as easy as <= 
  return !failure && ( l_F <= l ) // subset ...  
}
\end{lstlisting}

\end{subsection}

\begin{subsection}{Comparing Tabular Results}
Given two list of tuples $L,R$ ( perhaps results of two different versions of the same reporting software )
verify if they are indeed same list or not. The problem exists because the any tuple $t_l \in L$
can match to any tuple $t_r \in R$. The ordering of components in the tuples $t_l,t_r$ might differ. 
Hence, there is a precise ordering imposed by the ordering index sets $I_l,I_r$, 
such that for index $i$ component $t_l[I_l[i]]$ must be compared with component $t_r[I_r[i]]$. 
The ZoomBA solution is as follows:

\begin{lstlisting}[style=ZoomBAStyle]
def verify_tables(l,r, I_l, I_r){
  // list of strings from tuples `l' each has components ordered by I_l 
  ll = list(l) as {
    t = $.o ; str(I_l,'#') as { t[$.o]  }  }
  // list of strings from tuples `r' each has components ordered by I_r 
  lr = list{ // store the tuple
    t = $.o ; str(I_r,'#') as { t[$.o]  }  }
  // now compare ...
  ll == lr 
} 
\end{lstlisting}
\end{subsection}
\end{section}

\begin{section}{Summary \& Resources}

\begin{subsection}{Summary}
As almost all of modern enterprise application are written using JVM/CLR stack, 
it is impossible to avoid underlying runtime and write system integration/automation,
because in many cases one would need to call appropriate runtime methods to automate APIs. 
Being imperative does not help, because who tests the test code itself? 
These examples presented showcase how declarative
paradigm (almost SQL like) can be used to validate problems,
all of which would have otherwise required many lines of non-verifiable coding.
Similar problems exists for data manipulation/adaption layer. 

This is the idea behind ZoomBA : a declarative and functionally inclined open sourced language that incorporates all the good stuffs from the vast runtime libraries, while not being verbose enough to let system integrators focus on verifiable strategies, 
not on writing imperative code to solve a problem which can easily be solved declaratively. ZoomBA itself is an example of TDD 
in practice, having 80\% instruction coverage by unit tests. 
Bayestree uses ZoomBA for data adaption and system integration, and for validation problems of any nature.
Empirically, the learning curve for ZoomBA is found to be a from two weeks to a month.
\end{subsection}

\begin{subsection}{Online Resources}
The following are the online resources :
\begin{enumerate}
\item{\textbf{Download Location for binary for this Document \\}\url{https://gitlab.com/non.est.sacra/zoomba/blob/master/dist/beta5-snapshot.zip}}
\item{\textbf{Installation Instructions \\} \url{https://gitlab.com/non.est.sacra/zoomba#start-using-zoomba}}
\item{\textbf{Source Code } \url{https://gitlab.com/non.est.sacra/zoomba} }
\item{\textbf{Wiki Pages }\url{https://gitlab.com/non.est.sacra/zoomba/wiki}}
\end{enumerate}
\end{subsection}
\end{section}

\end{document}